\documentclass[amsmath,amssymb,
 aps,
 pra]{revtex4-2}
\usepackage{graphicx}
\usepackage{dcolumn}
\usepackage{bm}
\usepackage{physics}
\usepackage{float}

\begin{document}

\title{Impact of Jahn-Teller distortions on persistent molecular ring current in benzene}

\author{T. Joyce}
\author{A. Jaron}

\affiliation{
JILA and Department of Physics, University of Colorado, Boulder, CO-80309, USA
}

\begin{abstract}
Circularly polarized femtosecond UV laser pulse can remove a $\pi$ electron from benzene in such a way that the leftover hole circulates around the cation as a persistent ring current.
We investigate the time dependent strength of the current as the molecule relaxes from the $D_{6h}$ symmetry of the neutral to the $D_{2h}$ symmetry of the cation due to the Jahn-Teller effect.
We explore the effect of spontaneous symmetry breaking on persistent ring currents for benzene cation, because it is  one of the most comprehensively studied examples of the Jahn-Teller effect.

\end{abstract}

\maketitle

\section{Introduction}

With attosecond laser pulses, it is possible to directly measure and control electronic processes in molecules on their natural time scale 
. This technology is constantly improving, and a particularly significant example of recent breakthrough is the production of attosecond pulses  with tunable polarization 
. 

In addition to the
exciting applications that have already been demonstrated, circularly
polarized attosecond pulses may enable the study of photoinduced molecular ring currents,
a relatively longstanding theoretical prediction. The name implies an analogy to aromatic ring currents, a theoretical description in
chemistry where delocalized $\pi$-electrons found in aromatic molecules circulate freely in response
to an external magnetic field. Ring currents induced by ultrashort laser
pulses are predicted to have several distinct advantages over those induced by static magnetic fields: first, the
current is orders of magnitude stronger, as is the internal magnetic field; second, they enables
femtosecond (or even attosecond) time-resolved studies of aromaticity and magnetism; third,
they offer greater opportunity for coherent control of molecular currents via pulse shaping, which
may have applications for controlling chemical reactions or nanoscale electronic devices.
Because of the lack of experimental data, and because most previous theoretical studies

There has been much interest in ultrafast dynamics in aromatic molecules driven by circularly polarized laser pulses \cite{ulusoy2011, hermann2016, jia2017,mineo2017}.
In particular, a persistent ring current can be created when the laser excites the molecule into a superposition of either degenerate or quasidegenerate states \cite{barth2006,nobusada2007}.
In contrast to ordinary aromatic ring currents produced in response to an external magnetic field, these persistent ring currents are orders of magnitude stronger and do not require any external fields beyond the initial laser pulse.
The roles of molecular symmetry and vibrations have been explored by Kanno et al., who showed that ring currents in low-symmetry molecules reverse direction periodically and gradually decay due to nonadiabatic coupling \cite{kanno2010}.
Despite these results, it has generally been assumed that nuclear motion can be neglected for studies of  ultrafast electron dynamics in high-symmetry molecules such as benzene \cite{mineo2014}.

Recently we studied persistent ring currents induced in a benzene molecule (C$_6$H$_6$) after interacting with a circularly polarized ultrashort laser pulse \cite{tjaj2023}.
Starting from the same initial state, namely the  ground state for neutral benzene molecule, depending on the peak laser intensity, it is observed that the current can be predominantly carried by either electrons or holes. In general, hole currents are a consequence of helicity dependent resonance enhanced multiphoton ionization (REMPI), and electrons that orbit with or against the rotating laser polarization are ionized at different rates.
Using time-dependent density functional theory simulations, we demonstrated that helicity dependent ionization could be achieved and we visualized the appearance of two opposite types of persistent ring currents, of electrons and holes. 

In present paper we we consider the case how a circularly polarized femtosecond UV laser pulse can remove a $\pi$ electron from benzene in such a way that the leftover hole circulates around the cation in a persistent ring current.
In particular we investigate the time dependent strength of the persistent ring current as the molecule relaxes from the $D_{6h}$ symmetry of the neutral to the $D_{2h}$ symmetry of the cation due to the Jahn-Teller effect.
Let us note that this is also interesting starting point for exploring the effect of spontaneous symmetry breaking on persistent ring currents, because the benzene cation is one of the most comprehensively studied examples of the Jahn-Teller effect available.

This paper is structured as follows. In Sec. \ref{sec:model} we will discuss our time-dependent model for persistent currents and the Jahn-Teller effect in the benzene cation.
In Sec. \ref{sec:results} we discuss the results of our numerical simulations.
We end with a conclusion in Sec. \ref{sec:conclusion}.

\begin{figure}
    \includegraphics[width=0.50\linewidth]{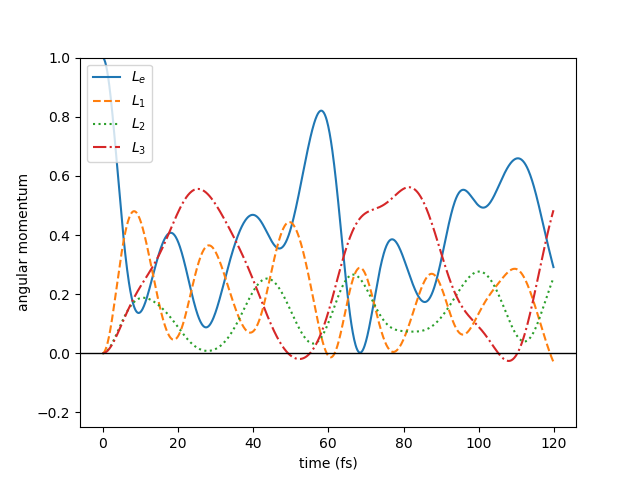}
    \includegraphics[width=0.50\linewidth]{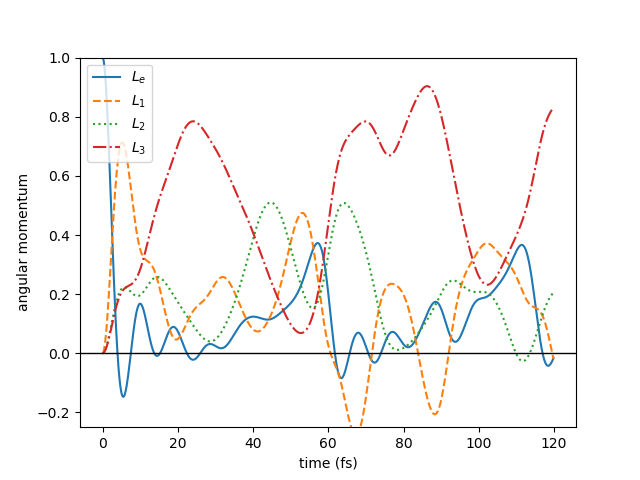}
    \includegraphics[width=0.50\linewidth]{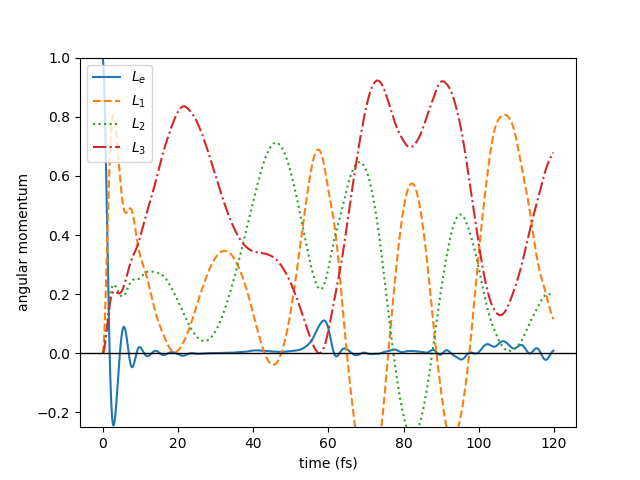}
    \caption{Redistribution of angular momentum among electronic and nuclear degrees of freedom as a function of time after sudden ionization of benzene by a circularly polarized UV laser pulse. The plots correspond to different coupling strengths: $\gamma=0.5$ (left), $\gamma=1$ (center), $\gamma=2$ (right). Physically correct parameters are $\gamma=1$.}
    \label{fig:angmom}
\end{figure}

\begin{figure}
    \centering
    \includegraphics[width=0.50\linewidth]{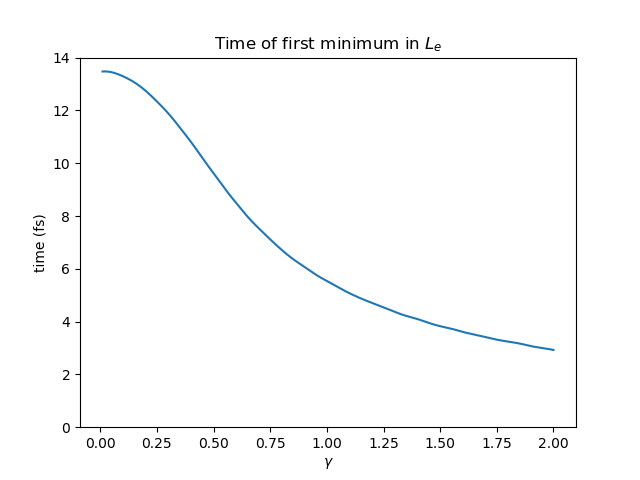}
    \caption{Time after ionization of the first minimum as a function of the coupling parameter $\gamma$.}
    \label{fig:firstMin}
\end{figure}

\begin{figure}
    \centering
    \includegraphics[width=0.50\linewidth]{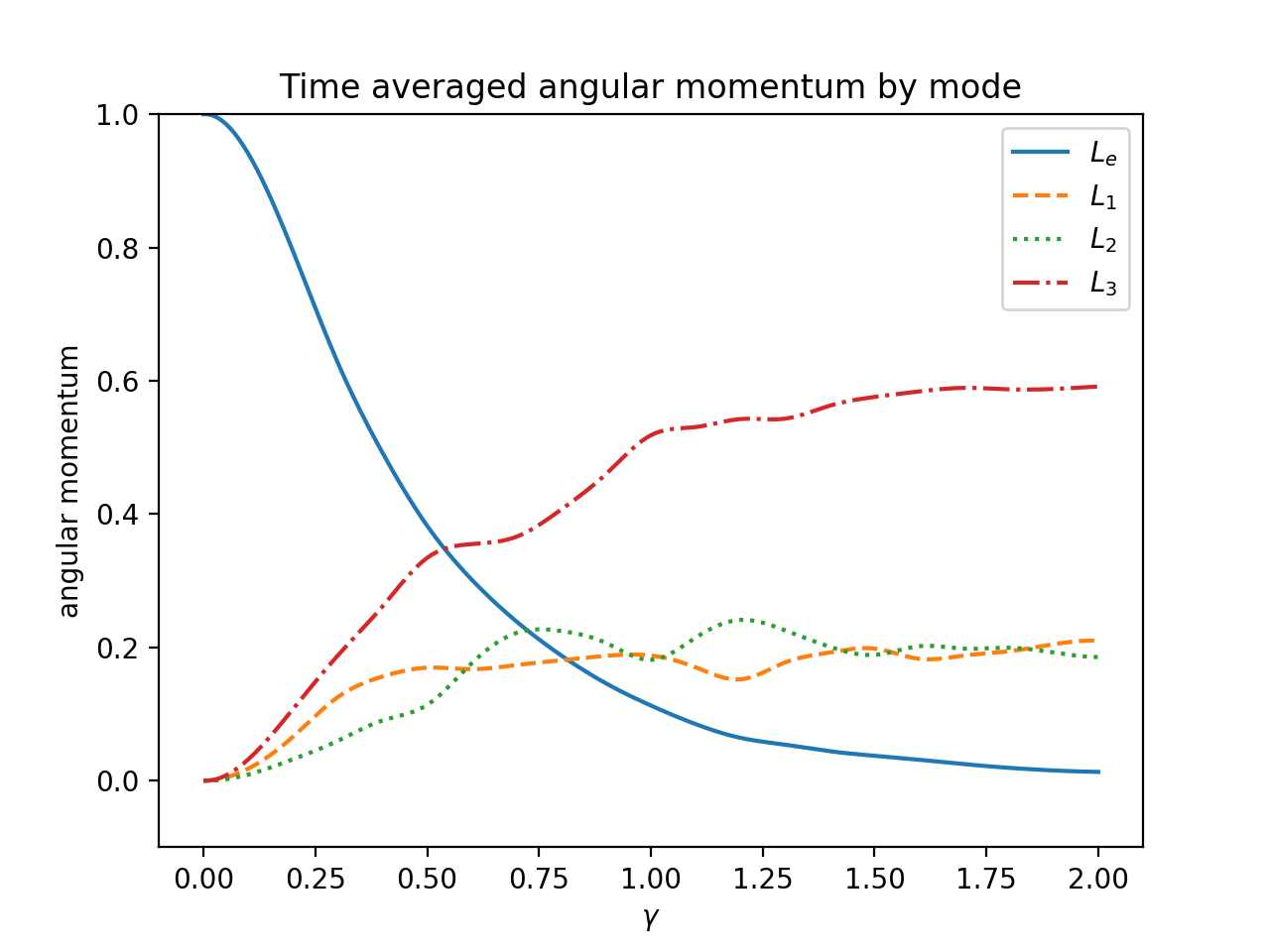}
    \caption{Time-averaged angular momentum in each mode, as a function of coupling strength.}
    \label{fig:average}
\end{figure}

\section{Model}
\label{sec:model}
The lowest two electronic states ($E_{1g}$) of the benzene cation are degenerate at the initial $D_{6h}$ geometry, and exhibit a conical intersection when the symmetry is broken.
We restrict our model to these lowest two potential energy surfaces, using the diabatic basis
\begin{align}
    \Psi_\pm = \Psi_1 \pm i \Psi_2
\end{align}
where $\Psi_{1,2}$ are an orthogonal pair of real-valued electronic eigenfunctions at the reference $D_{6h}$ configuration.
We model the vibronic Hamiltonian on the two diabatic surfaces for $\Psi_\pm$ as a harmonic potential plus linear Jahn-Teller coupling to three $e_{2g}$ vibrational modes:
\begin{align}
    \hat{H} = \sum_{n=1}^3 \omega_n \begin{bmatrix}
a_{n,+}^\dagger a_{n,+} + a_{n,-}^\dagger a_{n,-} & \gamma d_n(a_{n,-} + a_{n,+}^\dagger)\\
\gamma d_n(a_{n,+ } + a_{n,-}^\dagger) & a_{n,+}^\dagger a_{n,+} + a_{n,-}^\dagger a_{n,-}
\end{bmatrix}.
\end{align}
The operators $a_{n,\pm}^\dagger$ create phonons in the $n$th vibrational mode corresponding to clockwise or counterclockwise pseudorotations respectively.
We take the values of the parameters $\omega_n$ and $d_n$ from \cite{applegate2002}, where they were obtained from high-accuracy \textit{ab initio} electronic structure calculations, and were shown to compare favorably with measured vibronic spectra:
\begin{align}
 \nonumber
 &\omega_1 = 1571~ \text{cm}^{-1}, \quad d_1 = 0.68,\\
 \label{eq:parameters}
 & \omega_2 = 1152~ \text{cm}^{-1}, \quad d_2 = 0.49, \\
 \nonumber
 & \omega_3 = 573~ \text{cm}^{-1}, \quad d_3 = 0.92.
\end{align}
There is, in principle, a fourth $e_{2g}$ mode with $\omega_4 = 3017~\text{cm}^{-1}$, but the coupling parameter $d_4$ is small enough that it can be neglected.
No other modes are allowed by symmetry to be Jahn-Teller active at linear order.
The extra parameter $\gamma$ is adjustable so that we can control the strength of the Jahn-Teller coupling. The physically correct parameter values are given by $\gamma = 1$, but we will also investigate the impact of weaker ($\gamma<1$) or stronger ($\gamma>1$) coupling.
Relevant observables---such as the electric current around the ring, the magnetic moment, or any kind of circular dichroism signal---are all proportional to the expectation value of
\begin{align}
    \hat{L}_e &= \begin{bmatrix}
1 & 0\\
0 & -1
\end{bmatrix},
\end{align}
which is conveniently interpreted as electronic angular momentum about the molecular axis. 
The corresponding nuclear angular momentum in the $n$th vibrational mode is
\begin{align}
\hat{L}_n &= 
2\left(a^\dagger_{n,+} a_{n,+} - a^\dagger_{n,-} a_{n,-}\right)
\begin{bmatrix}
1 & 0\\
0 & 1
\end{bmatrix},
\end{align}
so that total angular momentum is conserved,
\begin{align}
    \left[\hat{H},\hat{L}_e + \sum_{n=1}^3 \hat{L}_n \right] = 0.
\end{align}

At time $t=0$, the electronic wavefunction is strictly $\Psi_+$, which is an eigenfunction of $\hat{L}_e$ with eigenvalue $+1$.
As time moves on, angular momentum transfered from the electrons to the nuclei via the Jahn-Teller coupling, which causes the strength of the electronic current to decrease correspondingly.
For simplicity, the initial nuclear wavepacket is taken to be the ground state of the harmonic potential (no phonons in any vibrational mode), although technically it should be the ground state of the neutral rather than the cation potential, which has slightly different vibrational frequencies and modes.

To calculate the time evolution of this initial state, we solve numerically the time-dependent Schr\"{o}dinger equation (TDSE) with the time-independent Hamiltonian $\hat{H}$.
The problem is discretized by limiting each mode to no more than 7 phonons, giving a Hilbert space of dimension $2^{19}$ in which $\hat{H}$ is represented as a sparse matrix.
We then implement Crank-Nicolson time propagation using sparse matrix operations from the Scipy library.
We have verified that our results are converged for a time step of $dt = 1$ a.u. and a maximum of 7 phonons per mode.

\section{Results}
\label{sec:results}

Results are shown in Fig. \ref{fig:angmom} for various coupling strengths $\gamma=0.5,1,2$.
In all cases, the one unit of angular momentum which initially belongs to the electrons is rapidly transferred to the first vibrational mode (yellow, dashed) in only a few femtoseconds, then later to the other two modes.
The third vibrational mode (red, dot-dashed), which has both the smallest frequency and the largest coupling constant, appears to hold most of the angular momentum on longer time scales.
There are revivals in the electronic angular momentum every $\sim 60$ fs, which corresponds to the vibrational period of the third mode.
The second vibrational mode has the lowest coupling constant, and appears to be the least important for these dynamics. Nonetheless, there are times when that mode holds a significant portion of the angular momentum, and therefore the problem necessitates a full three mode description.

Adjusting the coupling strengths $d_i$ appears to have two main effects on the dynamics. First, the time scale for the initial angular momentum transfer depends strongly on the coupling strength. This can be seen in Fig. \ref{fig:firstMin}, where the time after ionization of the first minimum in $L_e$ is plotted as a function of the coupling strength $\gamma$.
For weak coupling it takes over 10 fs for the Jahn-Teller distortion to affect the current, whereas for strong coupling it happens in fewer than 4 fs.

The second effect of the coupling strength is to decide the long term balance between electronic and nuclear angular momentum.
This is illustrated further in Fig. \ref{fig:average}, where the time-averaged angular momentum in each mode is compared as a function of coupling strength $\gamma$.
For each point on this plot, we solve the TDSE up to a maximium time $T=500$ fs, then compute the windowed averages,
\begin{align}
    \bar{L}_e &= \int_0^T \expval{\hat{L}_e}_t W(t/T) dt \\
    \bar{L}_n &= \int_0^T \expval{\hat{L}_n}_t W(t/T) dt,
\end{align}
where $W(t)$ is a Blackman window.

\section{Conclusion}
\label{sec:conclusion}
In conclusion, Jahn-Teller distortion should have a significant impact on persistent ring currents in the benzene cation. Although the current is not completely destroyed by these interactions, it is reduced to about 10\% of what the frozen nuclei model predicts.
There are partial revivals of the electronic current up to $40\%$ every $\sim 60$ fs, corresponding to the vibrational period of the $\nu_{18}$ mode (which has the strongest Jahn-Teller coupling strength).

The onset of the Jahn-Teller effect after ionization in benzene has been a topic of recent interest in itself \cite{tachikawa2018}.
We have shown in this paper that the time scale for the redistribution of angular momentum through Jahn-Teller coupling can be as short as 4 fs, because of its strongly nonadiabatic character,



{}
\end{document}